\documentclass[english]{article}
\usepackage[T1]{fontenc}
\usepackage[latin9]{inputenc}
\usepackage{geometry}
\usepackage{array}
\usepackage{float}
\usepackage{booktabs}
\usepackage{multirow}
\usepackage{amstext}
\usepackage{graphicx}
\usepackage{rotating}

\makeatletter

\newcommand{\lyxmathsym}[1]{\ifmmode\begingroup\def\b@ld{bold}
  \text{\ifx\math@version\b@ld\bfseries\fi#1}\endgroup\else#1\fi}

\providecommand{\tabularnewline}{\\}

\newcommand{\lyxaddress}[1]{
	\par {\raggedright #1
	\vspace{1.4em}
	\noindent\par}
}

\makeatother

\usepackage{babel}
\begin{document}
\title{First-principles predictions of tunable half metallicity in zigzag
GaN nanoribbons with possible applications in CO detection and spintronics}
\author{Rachana Yogi$^{1}$, Kamal K. Jha$^{2}$, Alok Shukla$^{1}$ and Neeraj
K. Jaiswal$^{*}$}
\maketitle

\lyxaddress{$^{1}$ Department of Physics, Indian Institute of Technology Bombay,
Powai, Mumbai 400076 India.}

\lyxaddress{$^{2}$ Indian Institute of Information Technology, Vadodara, Gujarat
382028 India.}

\lyxaddress{$^{*}$2D Materials Research Laboratory, Indian Institute of Information
Technology Design $\&$ Manufacturing, Jabalpur, M.P. 482005, India.}

E-mail: neeraj@iiitdmj.ac.in
\begin{abstract}
Based on systematic first-principles density-functional theory (DFT)
simulations, we predict that the zigzag GaN nanoribbons (ZGaNNR) can
be used both as highly efficient CO detectors as well as spin filters.
Our calculations, performed both on infinitely long nanoribbons, and
also on finite strands, suggest that: (a) CO binds strongly at the
edges of ZGaNNRs, and (b) that several of the resultant configurations
exhibit half-metallic behavior. We considered various edge-passivation
sites and found that all the resultant structures are thermodynamically
stable. The metallic, half-metallic, and semiconducting configurations
are observed as a function of CO passivation coverage. We also compute
the current-voltage (I-V) characteristics of various structures using
the Landauer formalism, and find that the devices made up of half-metallic
configurations act as highly-efficient spin filters. The effect of
CO concentration is also investigated which suggests a viable way
to not just tune the electronic band gap of ZGaNNRs, but also their
half metallicity. Our simulations thus suggest a new direction of
research for possible device applications of III-V heterostructures.
\end{abstract}
\begin{itemize}
\item \textbf{Keywords}: Nanoribbons, Gallium Nitride, Zigzag, Carbon monoxide,
detection/capturing, spin filtering.
\end{itemize}

\section*{Introduction}

2-D materials \cite{geim2010,biswas2007,Tsipas2013,ghosh2010} exhibit
potential candidature as a building block in upcoming sensing devices
due to their higher surface to volume ratio, unique electron confinement
and mature synthesis techniques compared to their 1-D counterparts
\cite{peyghan2014,yoon2011,samadizadeh2015,huang2008,monshi2017,yogi2019}.
Their outstanding electronic, magnetic and transport properties, higher
carrier mobility also makes them center of attraction for various
technological applications \cite{gonzalez2019,cui2020,xiao2017,chen138,guan2021,luo2021,sun2021,cui2021,guan2020}.
It is reported that nanoribbons can be realized either by cutting
\cite{hiura2004} mechanically exfoliated nanosheet \cite{maitra2012}
or by patterning epitaxially grown nanosheet \cite{berger2006,berger2004}.
The obtained nanoribbons are in the form of zigzag, armchair and chiral
\cite{jia2009,fujita1996,nakada1996}. Edge shape play a vital roll
in case of nanoribbons as all the electronic, magnetic and transport
properties are significantly influenced by their edge geometries \cite{tang2011,ezawa2006,ritter2009,li2010,tan2011,du2007,nakada1996}.
GaN nanoribbons are successfully synthesized using various technique
\cite{bae2002,yang2004,biswas2007,lo2003synthesis}. The electronic
and magnetic properties of nanoribbons make them a good candidate
for various applications \cite{zheng2011}.

Continuous advancement in industries and growth of traffic increases
the pollution rate day by day. Therefor, the development of highly
efficient sensors is one of the most active area of research nowadays.
The exceptional electronics properties and edge reactivity of 2D material
makes them a good candidate for detection of gas molecule at very
low concentration \cite{xia2021sensing,khan2017two,cui2020a,wu2021}.
GaN nanoribbons exhibit wide band gap and a promising material for
optoelectronic and high power applications. On the other hand, its
investigations towards sensing devices are very limited \cite{chen2019,yogi2019,yogi2020}.
The application of GaN in sensing devices could be superior compared
to graphene which is restricted via zero band gap behavior \cite{han2007,son2006}.
As CO is one of the most common environment pollutant gas, it is highly
warranted to develop CO sensing/capturing devices in near future.
On the other hand, the intrinsic wide band gap of GaNNR could obstruct
its path for semiconducting device applications. The interaction of
CO with GaNNR edges may affect their electronic properties. Therefore,
the present study, CO passivation with GaN based zigzag nanoribbons
(ZGaNNR) is investigated towards possible CO detection and its effect
on the band gap modulation. In our study we focused on the electronic
behavior of the material through which we have also analyzed the charge
transfer due to the passivation of gas molecule on ZGaNNR. Typically,
the charge transfer to the molecule or from the molecule, also induces
change in electric resistivity of the material.

\section*{Model and Methodology}

\begin{figure}[H]
\begin{centering}
\includegraphics[scale=0.4]{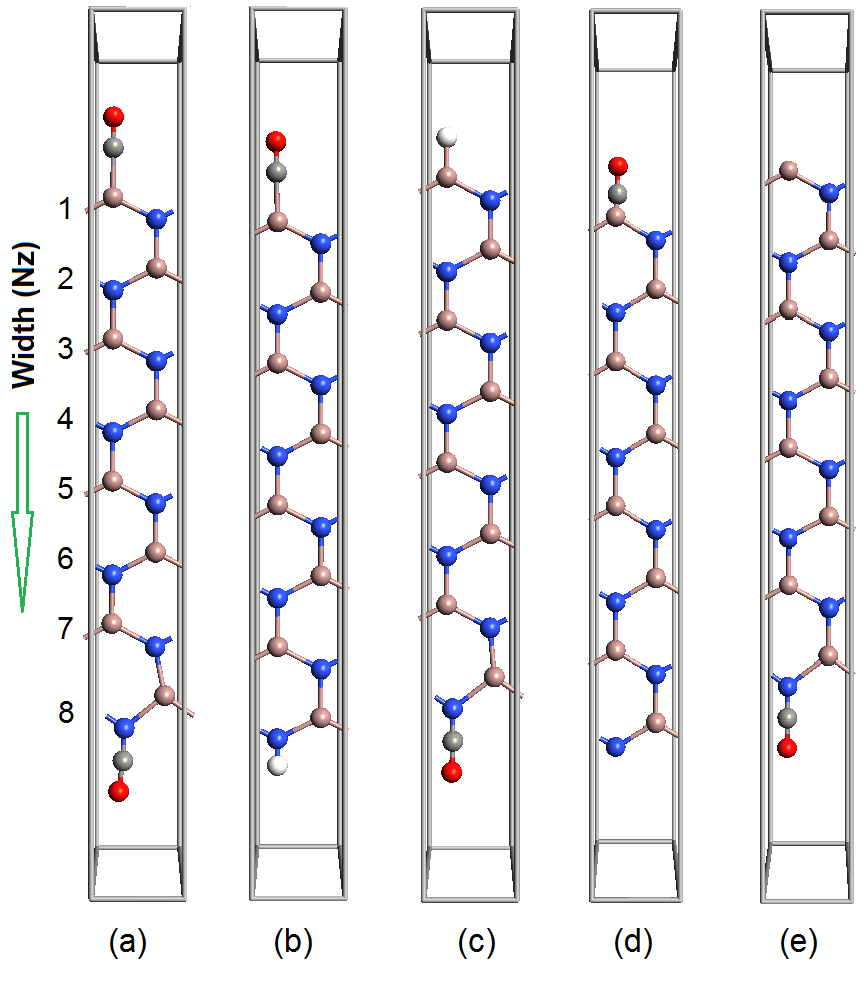}
\par\end{centering}
\caption{\label{fig:configuration}The schematics of zigzag GaN nanoribbon
with passivation of CO on (a) CO-ZGaN-CO, (b) CO-ZGaN-H, (c) H-ZGaN-CO,
(d) CO-ZGaN and (e) ZGaN-CO at width-8.}
\end{figure}

The present investigations were performed using first-principles self-consistent
calculations based on density functional theory (DFT). We used QuantumATK
DFT code \cite{Brandbyge2002} for the present simulation results.
The generalized gradient approximation (GGA) in the form of Predew-Burke-Ernzerhof
(PBE) was used as exchange correlation potential \cite{Perdew1996}.
The considered configurations of ZGaNNR were modelled with periodic
boundary conditions having repetition along Z-axis whereas X and Y
directions were kept confined. The norm-conserving pseudo-potential
with 70 Ry energy mesh cut-off value was adopted to define the fineness
of the grid. Further, double $\zeta$ polarized basis set has been
considered for all constituent atoms. The Monkhorst-Pack Grid \cite{Monkhorst1976}
for k-point sampling was selected as 1$\times$1$\times$50 for defining
the sampling of the Brillouin zone centered at $\Gamma$. To avoid
inter-ribbon interactions, ribbons were separated using a cell padding
vacuum of 10${\lyxmathsym{\AA}}$ along the confined directions. During
the geometry optimization, we used Puley mixer algorithm with 1$\times$10$^{-5}$Ry
tolerance for self-consistent iteration loop and all the atoms were
free to change their positions during optimization to attain the minimum
energy configuration. The geometries were relaxed without any constraint
till the force and stress on each atom reduces to a criterion of 0.05
eV/${\lyxmathsym{\AA}}$ and 0.05 eV/${\lyxmathsym{\AA}}$$^{3}$
respectively. The ribbon width is defined in the conventional manner
$\emph{i.e.}$ the number of Ga-N bonds across the transverse direction
\cite{song2014, srivastava2015}. Five different passivation sites
on ZGaNNR are considered for CO molecule as illustrated in Fig. \ref{fig:configuration}.

\section*{Results and discussion}

\subsection*{Structural Stability}

To find out the most suitable passivation sites, various possible
configurations have been modeled. It is noticed that bonding of CO
molecule with the GaN nanoribbons takes place via C side as shown
in Fig. \ref{fig:configuration}. The variation in bond length and
bond angle has been illustrated in Table 1. To avoid an ambiguity,
the structural stability of CO passivated GaN nanoribbons is further
discussed in separate subsections for ZGaNNR.

\begin{table}
\caption{Calculated bond lengths and bond angles of CO passivated zigzag GaN
nanoribbons}

\centering{}%
\begin{tabular}{cc}
\hline 
\multicolumn{2}{c}{ZGaNNR}\tabularnewline
\hline 
Bond length ($\lyxmathsym{\AA}$) & Bond angle ($^{o}$)\tabularnewline
\hline 
C-O = 1.81 (Ga edge) & $\angle CGaN$ = 116.39 (Ga edge)\tabularnewline
C-O = 1.19 (N-edge) & $\angle GaNGa$ = 116.87 (Ga edge)\tabularnewline
C-Ga = 1.90 & $\angle HGaN$ = 118.08 (Ga edge)\tabularnewline
Ga-N = 1.86 & $\angle CNGa$ = 131.20-126.58 (N-edge)\tabularnewline
Ga-H = 1.53 & $\angle NGaN$ = 118.10 (N-edge)\tabularnewline
C-N = 1.25 & $\angle HNGa$ = 115.73 (N-edge)\tabularnewline
N-H = 1.25 & \tabularnewline
\hline 
\end{tabular}
\end{table}

The reported lattice constant of zigzag GaN nanoribbon is 3.18${\lyxmathsym{\AA}}$
which shows good agrement with our results (3.32${\lyxmathsym{\AA}}$)\cite{Sahin2009}.
All the properties of nanoribbons are highly dependent on their edge
shape and ribbon width. First we investigate the structural stability
of bare and H-passivated nanoribbons.

\begin{table}[H]
\caption{Variation of adsorption energy ($E_{ad}$), binding energy (BE), band
gap ($E_{g}$ ) and Fermi energy ($E_{F}$ ) of H-passivated and bare
zigzag GaN nanoribbons as a function of ribbon width.}

\begin{centering}
\begin{tabular}{ccccccc}
\toprule 
\multirow{3}{*}{Width ($N_{Z}$)} & \multicolumn{6}{c}{ZGaNNR (eV)}\tabularnewline
\cmidrule{2-7} \cmidrule{3-7} \cmidrule{4-7} \cmidrule{5-7} \cmidrule{6-7} \cmidrule{7-7} 
 & \multicolumn{3}{c}{H-Passivated} & \multicolumn{3}{c}{Bare}\tabularnewline
\cmidrule{2-7} \cmidrule{3-7} \cmidrule{4-7} \cmidrule{5-7} \cmidrule{6-7} \cmidrule{7-7} 
 & $E_{ad}$ & $E_{g}$ & $E_{F}$ & $BE$ & $E_{g}$ & $E_{F}$\tabularnewline
\midrule 
4 & -5.65 & 3.40 & -3.17 & -5.83 & M & -4.96\tabularnewline
\midrule 
6 & -5.69 & 3.04 & -3.55 & -6.02 & M & \multirow{1}{*}{-4.99}\tabularnewline
\midrule 
8 & -5.70 & 2.86 & -3.68 & -6.12 & M & \multirow{1}{*}{-4.98}\tabularnewline
\bottomrule
\end{tabular}
\par\end{centering}
\end{table}

To insure the structural stability we have calculated the binding
energy (BE) for bare ribbons and adsorption energy for H-passivated
ribbons respectively and the calculated values are depicted in Table
2. To calculate the BE following relation has been used \cite{prasongkit2019,loh2014}: 

\begin{equation}
BE=\frac{1}{m+n}[E_{total}-m(E_{Ga})-n(E_{N})]
\end{equation}

where E$_{total}$, E$_{Ga}$ and E$_{N}$ are, the total energies
of bare ribbon, isolated Ga, and N atom. Similarly, m and n represented
the number of Ga and N atoms in the nanoribbon respectively. As per
the definition adopted here, negative binding or adsorption energy
exhibits exothermic nature while the magnitude signifies thermodynamic
stability. It is noticed that structural stability increases with
the ribbon width. To find the most stable CO passivated nanoribbon,
we calculate adsorption energy ($\emph{E}$$_{ad}$) of considered
configurations including H passivation. The following relation has
been utilized for $\emph{E}$$_{ad}$ calculation: \cite{Sun2017,kumar2020} 

\begin{equation}
\emph{E}_{ad}=\frac{1}{n}[E_{T}-E_{bare}-nE_{H/CO}]
\end{equation}

where E$_{T}$, E$_{bare}$, E$_{H/CO}$ are the total energies of
considered configuration after attachment of CO/H, bare nanoribbon,
isolated CO molecule/H atoms, respectively and n is number of passivated
molecules/atoms. 

\begin{table}[H]
\caption{Variation of adsorption energy ($E_{ad}$), Fermi energy ($E_{F}$)
and band gap ($E_{g}$) of zigzag GaN nanoribbons with CO passivation
as a function of ribbon width.}

\centering{}%
\begin{tabular}{cccccc}
\hline 
\multirow{2}{*}{Configurations} & \multirow{2}{*}{Width ($N_{Z}$)} & \multirow{2}{*}{$E_{ad}$(eV)} & \multirow{2}{*}{$E_{F}$(eV)} & \multicolumn{2}{c}{Eg (eV)}\tabularnewline
\cline{5-6} \cline{6-6} 
 &  &  &  & Spin up & Spin down\tabularnewline
\hline 
\multirow{3}{*}{CO-ZGaN-CO} & 4 & -1.64 & -3.44 & M & 0.5\tabularnewline
 & 6 & -1.66 & -3.41 & M & 0.5\tabularnewline
 & 8 & -1.71 & -3.51 & M & 0.6\tabularnewline
\multirow{3}{*}{CO-ZGaN-H} & 4 & -3.67 & -3.06 & M & 3.0\tabularnewline
 & 6 & -3.71 & -3.05 & M & 2.6\tabularnewline
 & 8 & -3.71 & -3.04 & M & 2.5\tabularnewline
\multirow{3}{*}{H-ZGaN-CO} & 4 & -3.64 & -3.43 & 2.5 & 1.0\tabularnewline
 & 6 & -3.68 & -3.53 & 2.1 & 0.9\tabularnewline
 & 8 & -3.69 & -3.64 & 2.0 & 0.9\tabularnewline
\multirow{3}{*}{CO-ZGaN} & 4 & -1.19 & -4.49 & M & M\tabularnewline
 & 6 & -1.20 & -4.44 & M & M\tabularnewline
 & 8 & -1.75 & -4.81 & M & M\tabularnewline
\multirow{3}{*}{ZGaN-CO} & 4 & -2.65 & -3.51 & 1.0 & 0.2\tabularnewline
 & 6 & -2.69 & -3.55 & 0.9 & 0.2\tabularnewline
 & 8 & -2.68 & -3.63 & 0.8 & 0.2\tabularnewline
\hline 
\end{tabular}
\end{table}

For sensing of CO molecules, five different passivation sites are
considered as illustrated in Fig. \ref{fig:configuration}. The structural
stability of all considered geometries of ZGaNNR are tabulated in
Table 3. The magnitude of adsorption energy is much higher than thermal
excitation energy ($\approx$25 meV) which suggests that all the configurations
are thermally stable. Interestingly, it is also noticed that the most
energetically favorable configuration is CO-ZGaN-H followed by H-ZGaN-CO
from which we conclude that the presence of passivating H atoms effectively
increases the stability of ZGaN nanoribbon. Similarly, the order of
least favorable configurations are CO-ZGaN, CO-ZGaN-CO and ZGaN-CO.
The Fermi energy also varies as the ribbon width increases. In all
the configurations the Fermi level shifts in downward direction with
respect to its bare counterpart. The downward shifting of Fermi level
is analogous to p-type doping candidature. It is also concluded that
after passivation of CO, ZGaN nanoribbons are stable in antiferromagnetic
(AFM) ground state. We tried to incorporate the effect of vdW corrections
however, there was no change in the ground state energy of the structures.
This is due to the reason that all the adsorbed CO molecules form
stable chemical bonds with the host nanoribbons. Therefore, the weak
physical interactions are not affecting the ground state of the system
\cite{reckien2012, ilawe2015}. 

\subsection*{Electronic properties}

\begin{figure}[H]
\begin{centering}
\includegraphics[scale=0.35]{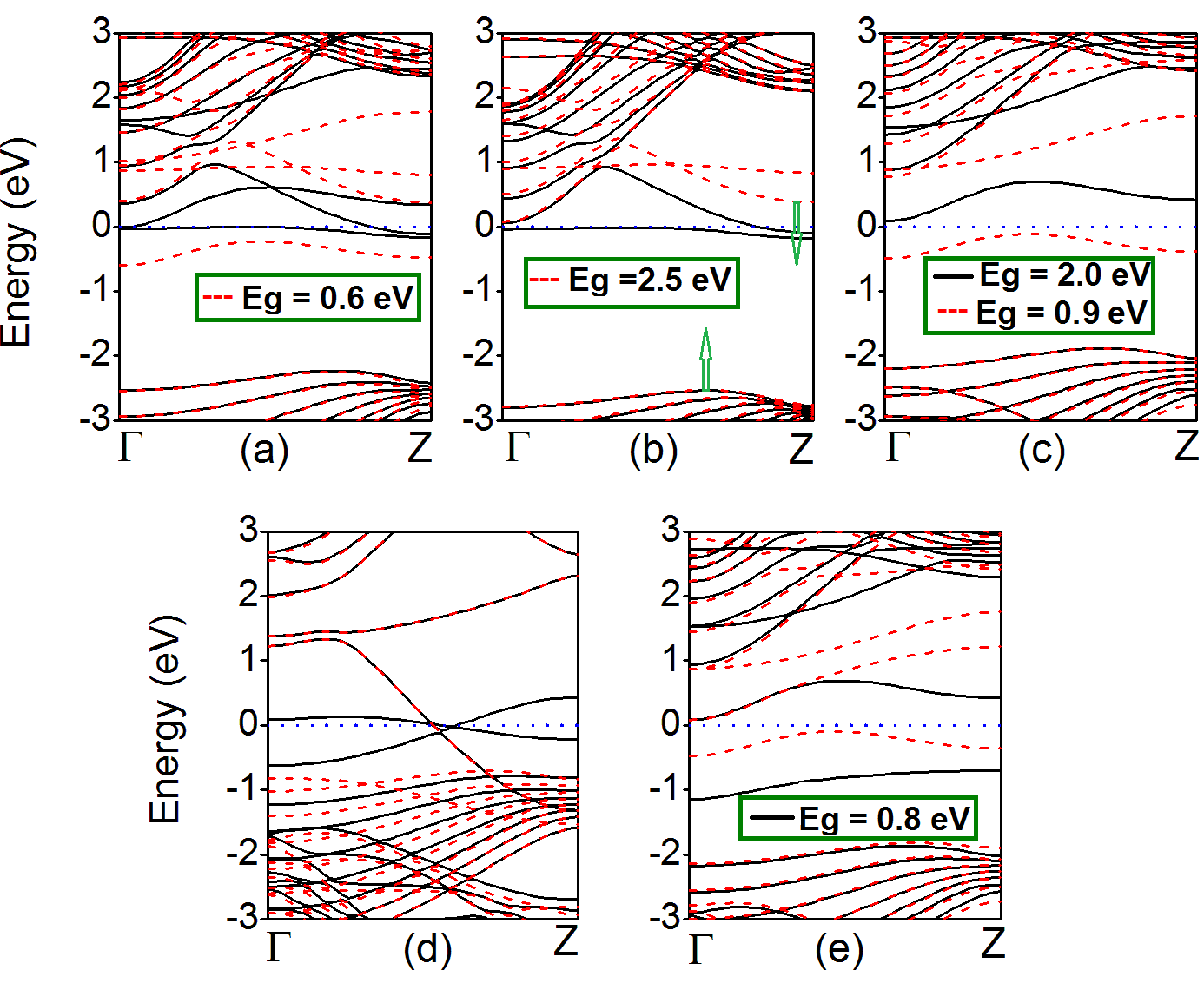}
\par\end{centering}
\caption{\label{fig:All-BS}The spin polarized band structure of zigzag GaNNR
with CO passivation for (a) CO-ZGaNNR-CO, (b) CO-ZGaNNR-H, (c) H-ZGaNNR-CO,
(d) CO-ZGaNNR and (e) ZGaNNR-CO at width-8. The solid (black) and
dashed (red) lines correspond to electronic states of spin up (majority
spin) and spin down (minority spin) electrons respectively. The horizontal
dotted line at 0 eV represents the Fermi level.}
\end{figure}

The electronic band structures of ZGaNNR with passivation of CO on
the edges of nanoribbon are illustrated in Fig. \ref{fig:All-BS}.
It is reported that zigzag bare GaNNR is metallic in nature whereas
H-passivation exhibits a band gap \cite{yogi2020,dai2012}. Also,
zigzag bare and H-passivated GaNNR, both are stable in non-magnetic
ground states, respectively. It is revealed that presence of CO on
the edges of ZGaNNR profoundly affects the electronic properties of
ZGaNNR. After passivation of CO, ZGaNNR is stable in AFM ground state.
The observed AFM ground state could be further explained on the basis
of electronegativity difference between C and the host edge atoms
(Ga and N). The electronegativity of C is greater (smaller) than the
Ga (N) atoms. Owing to this, an unequal charge transfer takes place
between the adsorbed CO molecule and the edge Ga/N atoms. It creates
a difference between the electric potentials of the opposite edges
which in turn favors the AFM ground state. Similar magnetic behavior
has been previously reported for graphene nanoribbons functionalized
with different atoms/groups \cite{jaiswal2017}. Interestingly, when
both edges of nanoribbon are containing CO molecule {[}Fig. \ref{fig:configuration}{]}
and when CO is passivated on the Ga edge and N edge posses H atom
{[}Fig. \ref{fig:configuration}(b){]} then half metallic nature is
noticed as depicted in Fig. \ref{fig:All-BS} (b). Another important
thing we have observed is the pure metallic character when CO is located
at the Ga edge of the ribbon {[}Fig. \ref{fig:All-BS} (d){]}. However,
when Ga edge is passivated by H atom or left bare, finite band gap
is observed {[}Fig. \ref{fig:All-BS} (c) and (e){]}. The calculated
band gap is 2.0 eV for spin-up carrier whereas 0.9 eV for spin-down
charge carries of H-ZGaN-CO configuration at ribbon width-8. Similarly,
ZGaN-CO configuration posses band gap {[}0.8 eV (spin-up) and 0.2
eV (spin-down){]}. It is reported that H-passivated ZGaNNRs exhibit
indirect band gap \cite{dai2012} and after passivation of CO the
behavior remains same {[}Fig. \ref{fig:All-BS} (c) and (e){]}.

\begin{figure}[H]
\begin{centering}
\includegraphics[scale=0.3]{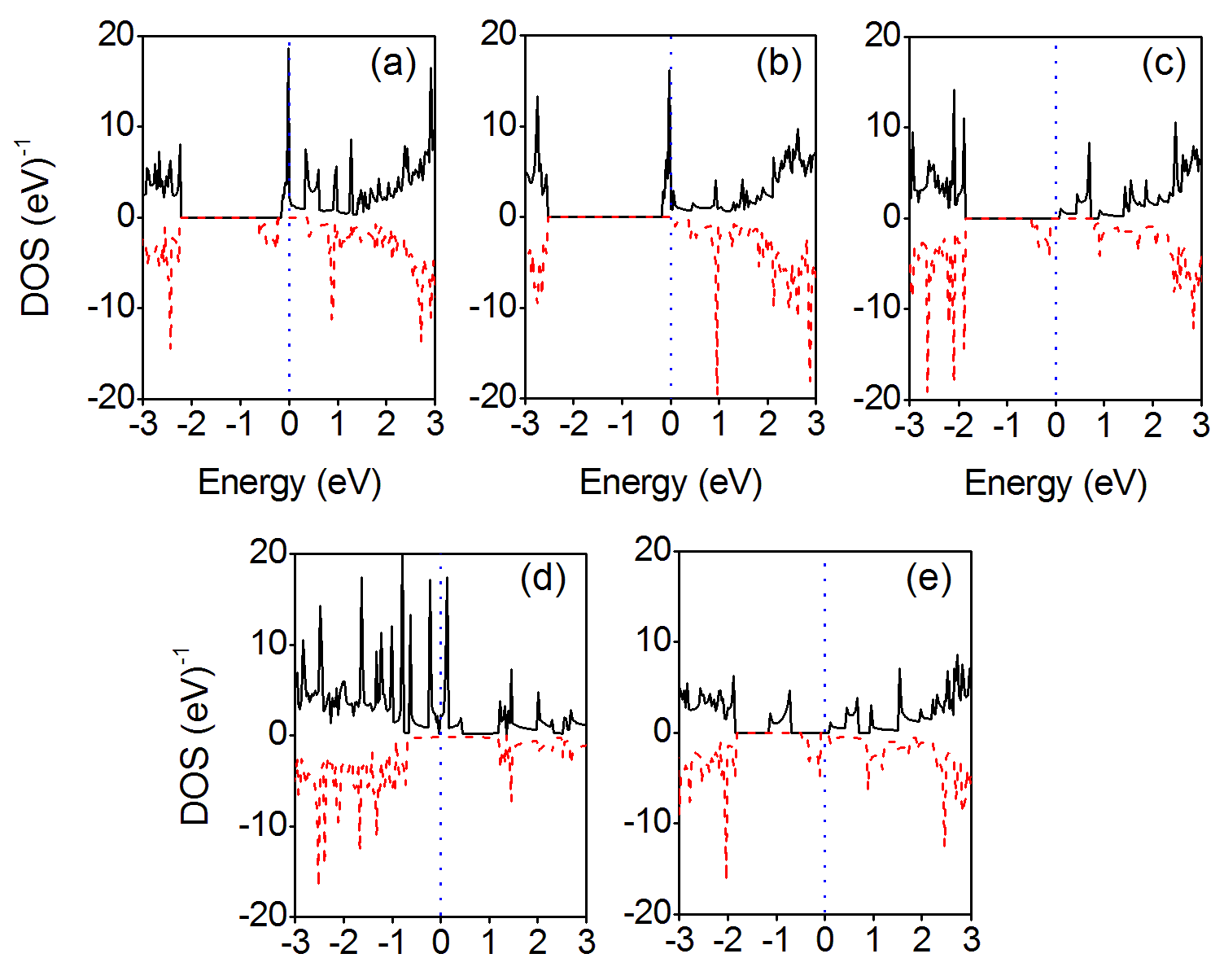}
\par\end{centering}
\caption{\label{fig:All-DOS}The spin polarized DOS of zigzag GaNNR with CO
passivation for (a) CO-ZGaNNR-CO, (b) CO-ZGaNNR-H, (c) H-ZGaNNR-CO,
(d) CO-ZGaNNR and (e) ZGaNNR-CO at width-8. The solid (black) and
dashed (red) lines correspond to electronic states of spin up (majority
spin) and spin down (minority spin) electrons respectively.}
\end{figure}

For the further understanding of electronic properties of ZGaNNR with
CO passivation, we have also calculated the density of states (DOS)
of all the considered geometries {[}Fig. \ref{fig:All-DOS}{]}. It
is revealed that the DOS profile exhibits good compatibility with
band structure results. The similar half metallic character is also
noticed in DOS of CO-ZGaN-CO and CO-ZGaN-H configurations. Similarly,
the absence of energy states near the Fermi level is witnessed in
H-ZGaN-CO and ZGaN-CO configuration, whereas the DOS of CO-ZGaN clearly
exhibits metallic behavior. 

\begin{figure}[H]
\begin{centering}
\includegraphics[scale=0.2]{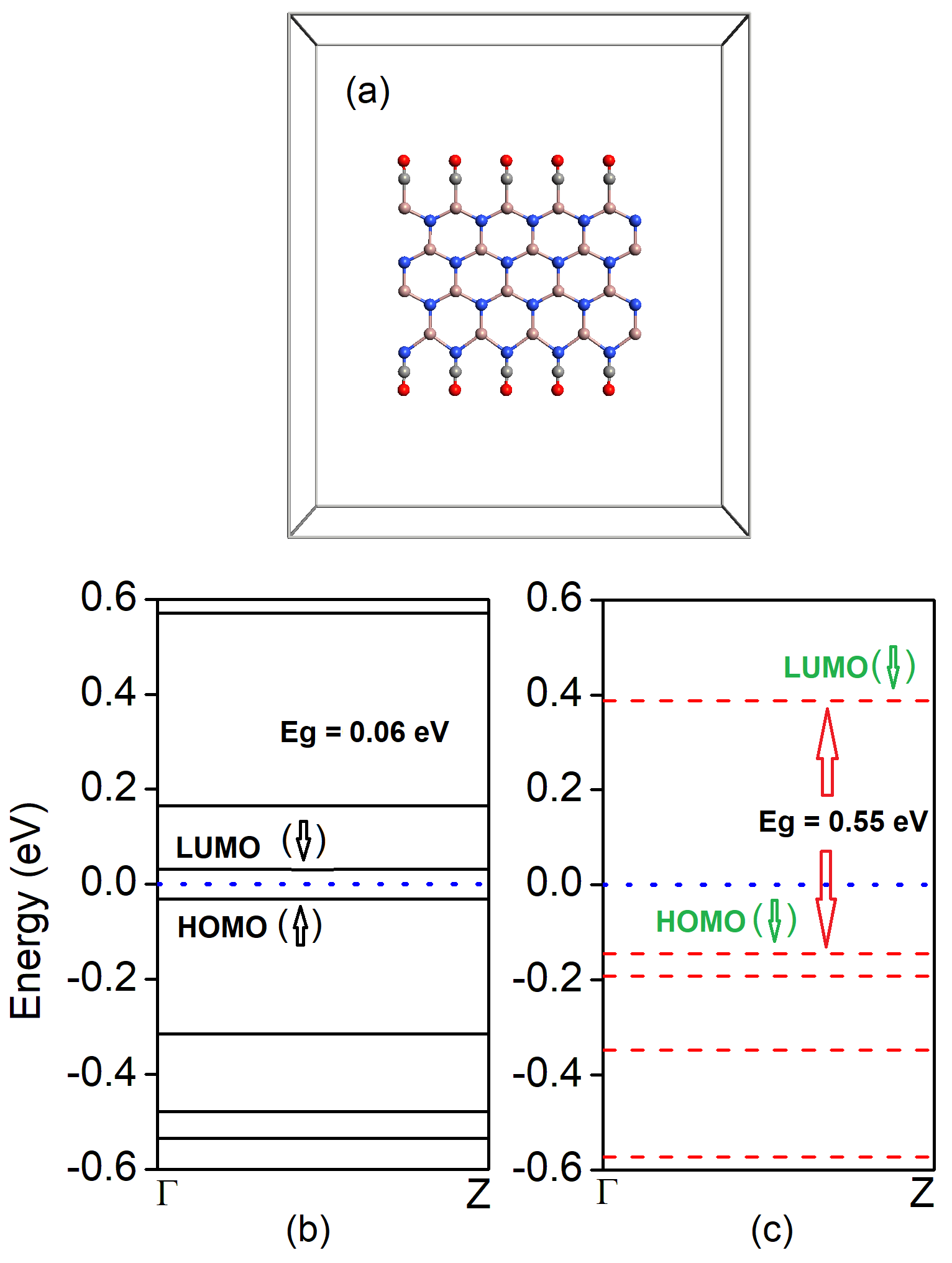}
\par\end{centering}
\caption{\label{fig:HOMO-LUMO}Energy levels of up- (b) and down-spin (c) electrons
for the CO-ZGaNNR- CO fragment consisting of five repeat units (a).
The solid (black) and dashed (red) lines correspond to electronic
states of spin up (majority spin) and spin down (minority spin) electrons
respectively. The blue dotted line at 0 eV represents the Fermi level.}

\end{figure}

In realistic devices, there will be finite fragments of nanoribbons
between the electrodes, unlike the infinitely long structures considered
for computing the band structure. Therefore, we have also investigated
the electronic properties of finite fragments of CO-ZGaNNR-CO consisting
of five unit cells (repetitions) as illustrated in Fig.\ref{fig:HOMO-LUMO}.
Our calculations reveal that even this finite fragment exhibits half-metallic
character with a finite band gap for spin down electrons ($E_{g}$=
0.55eV), and a negligible one ($E_{g}$= 0.06 eV) for spin up electrons.
Hence, only the spin-up electrons will participate in conduction,
leading to a spin-polarized current.

\begin{figure}[H]
\begin{centering}
\includegraphics[scale=0.5]{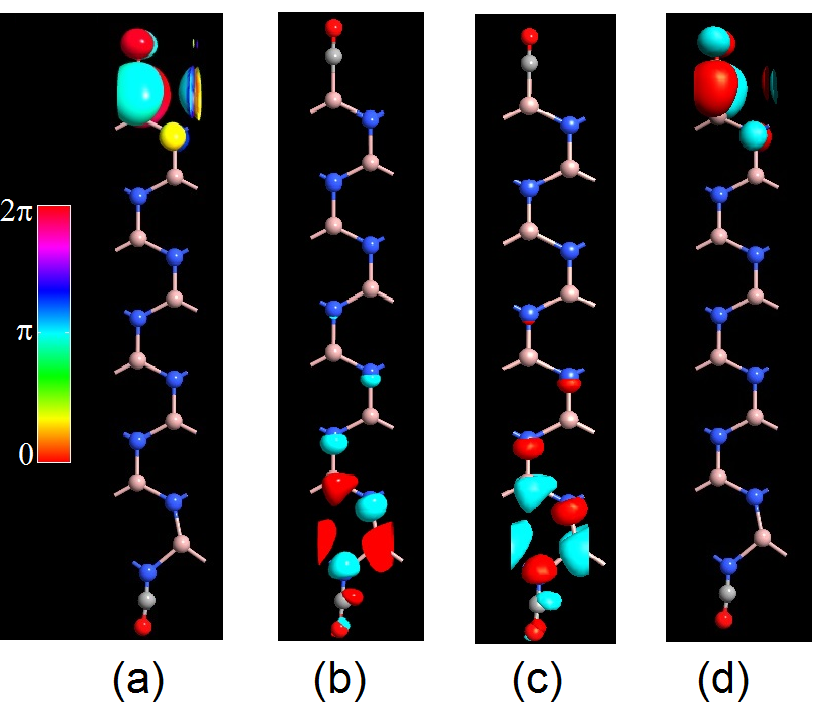}
\par\end{centering}
\caption{\label{fig:Bloch}The computed Bloch states for highest valence band
(HVB) and the lowest conduction band (LCB), plotted along the width
for (a) HVB-up spin (b) HVB-down spin (c) LCB-up spin and (d) LCB-down
spin.}

\end{figure}

For further verification of the observed half-metallic character and
understanding the splitting of electronic states around the Fermi
level, Bloch states have been analysed. We elected the highest valence
band and the lowest conduction band for this (Bloch) analysis as these
are mainly responsible for governing the electronic transport in the
material. The spin polarized Bloch states for CO-ZGaNNR-CO are depicted
in Fig. \ref{fig:Bloch}. Perusal of this figure reveals that different
spins are populated at opposite edges of the ribbon. The localized
behavior of electrons is also be noticed in Fig. \ref{fig:Bloch}
(a)-(d). Additionally, there exists a phase change of $\pi$ between
spatially separated charges on the two edges, further confirming the
AFM ordering in the system.

\subsection*{Transport Properties}

\begin{figure}[H]
\begin{centering}
\includegraphics[scale=0.3]{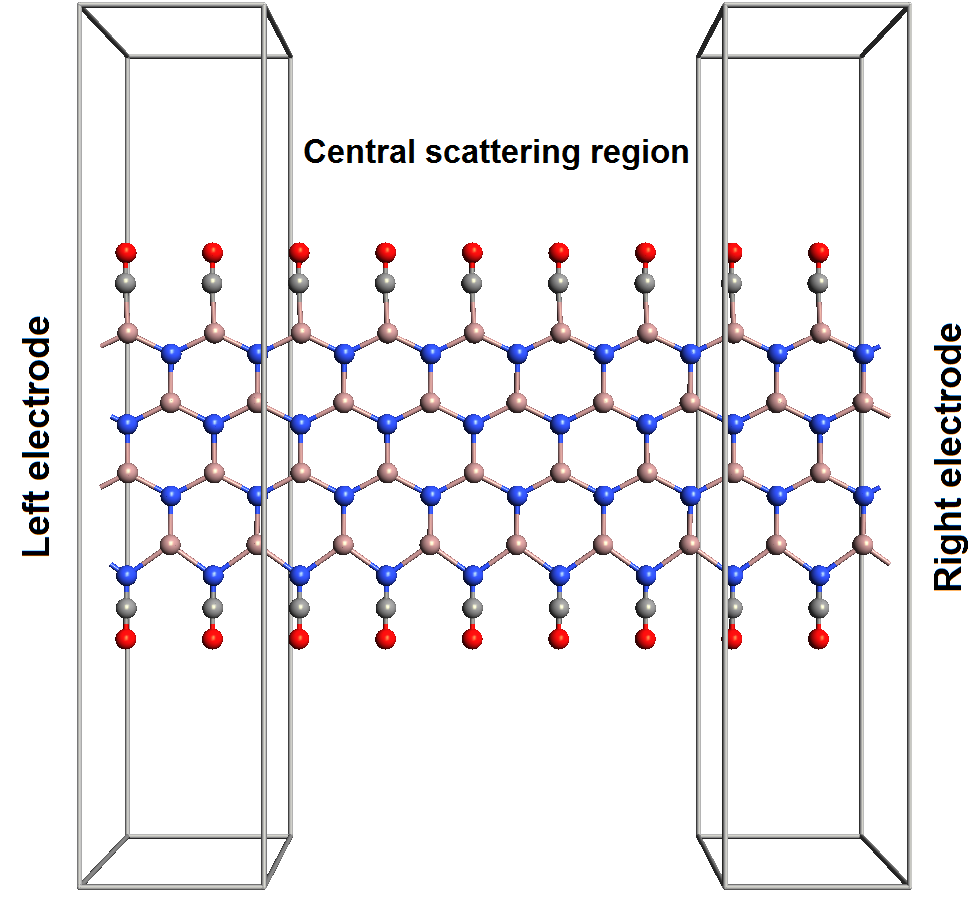}
\par\end{centering}
\caption{\label{fig:Two-probe}The Schematic diagram of the two probe geometry
of ZGaN nanoribbon with CO passivation for CO-ZGaN-CO at width-5 and
repetition 5 along Z axis. The nature of the device is identical for
other structures (Fig 1).}

\end{figure}

\begin{figure}[H]
\begin{centering}
\includegraphics[scale=0.3]{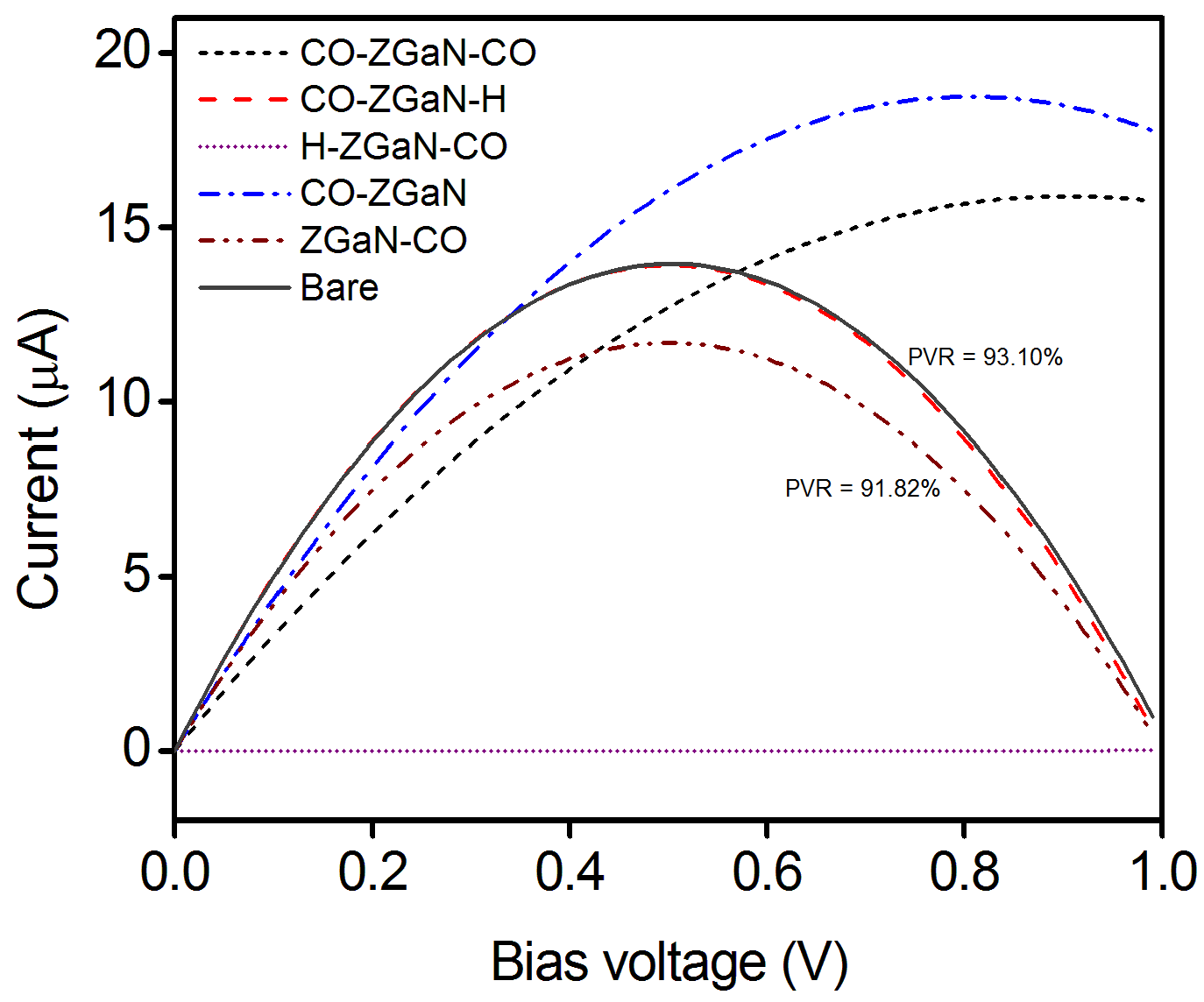}
\par\end{centering}
\caption{\label{fig:All-IV}The calculated I-V characteristic of bare and CO
passivated ZGaN nanoribbons with higher peak to valley ratio (PVR).}

\end{figure}

The two-probe model is used for transport \cite{chakrabarty2015}
studies as shown in Fig \ref{fig:Two-probe}. The calculated I-V characteristic
of ZGaNNR containing CO are illustrated in Fig \ref{fig:All-IV}.

It is revealed that CO-ZGaN configuration exhibits maximum current
followed by CO-ZGaN-CO. In these two structures, current increases
linearly upto $\sim$0.6 V beyond which it starts to saturate. In
contrast, an interesting behavior is noticed for rests of the other
I-V characteristics. For H-ZGaN-CO, the current remains zero for entire
bias window as the corresponding band gap is significantly larger
(2.5 eV) than applied biasing. For remaining three structures, the
current initially increases and attains a maximum value ($\sim$14
$\mu A$ for bare/CO-ZGaN-H and $\sim$ 11.6 $\mu A$ for ZGaN-CO)
around 0.5 V. As the biasing is further increased, the current starts
to decline and approaches 0 near 1 V. Thus, a clear signature of negative
differential resistance (NDR) phenomena is obtained in these three
structures.

\begin{figure}[H]
\begin{centering}
\includegraphics[scale=0.3]{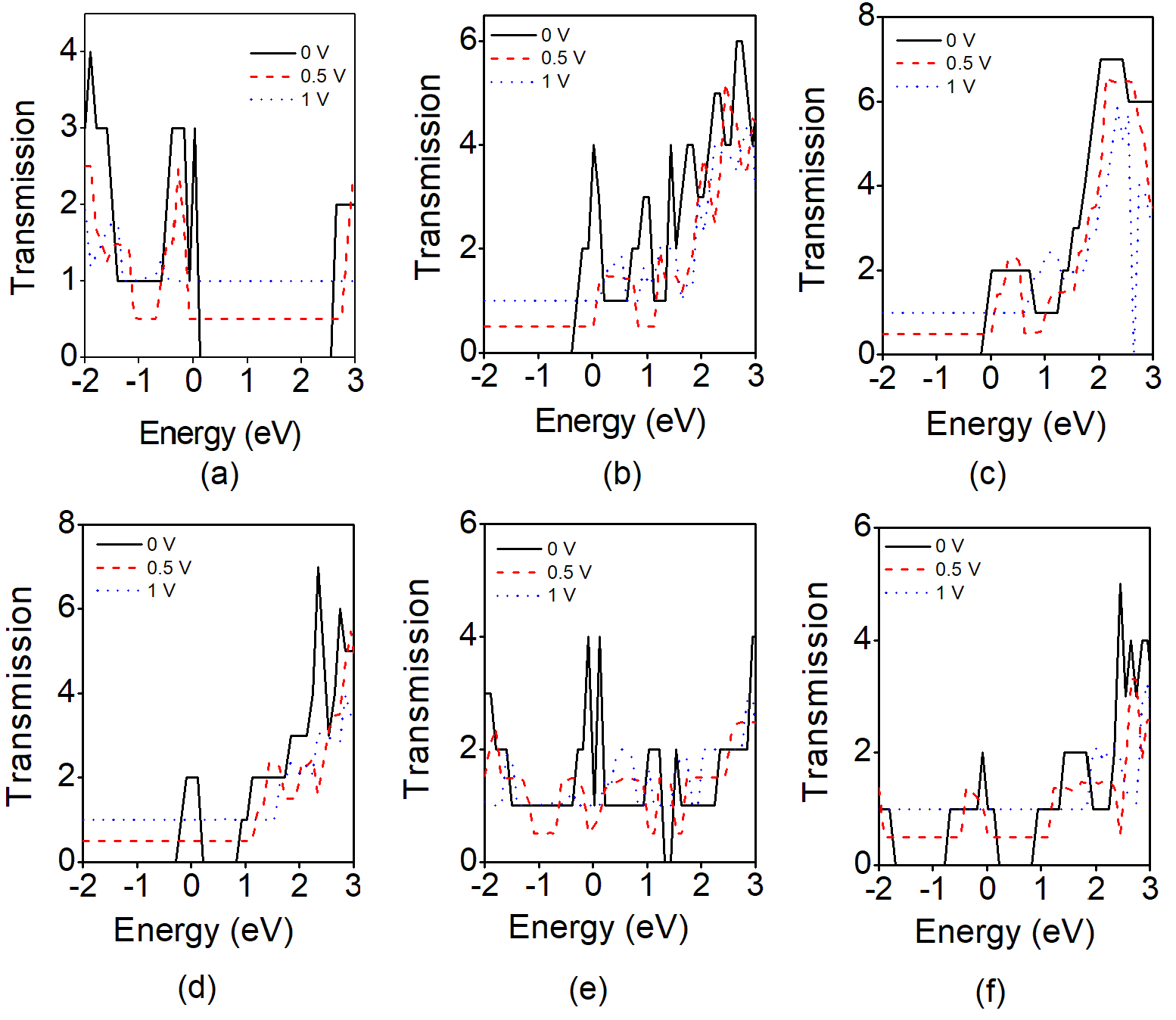}
\par\end{centering}
\caption{\label{fig:All-TS}The calculated transmission spectra of CO passivated
ZGaN nanoribbons for (a) bare, (b) CO-ZGaN-CO (c) CO-ZGaN-H, (d) H-ZGaN-CO,
(e) CO-ZGaN and (f) ZGaN-CO.}

\end{figure}

The observed I-V characteristics could be further understood on the
basis of transmission spectra as shown in Fig \ref{fig:All-TS}. For
CO-ZGaN and CO-ZGaN-CO, the transmission coefficient is found to be
maximum ($\emph{i.e.}$ 4) which supports highest current in these
structures. For the structures showing NDR behavior, a relatively
small but finite transmission coefficient is noticed around the Fermi
level. Upon increasing the biasing, reduction in the transmission
takes place which is responsible for the observed NDR phenomenon.

\begin{figure}[H]
\begin{centering}
\includegraphics[scale=0.3]{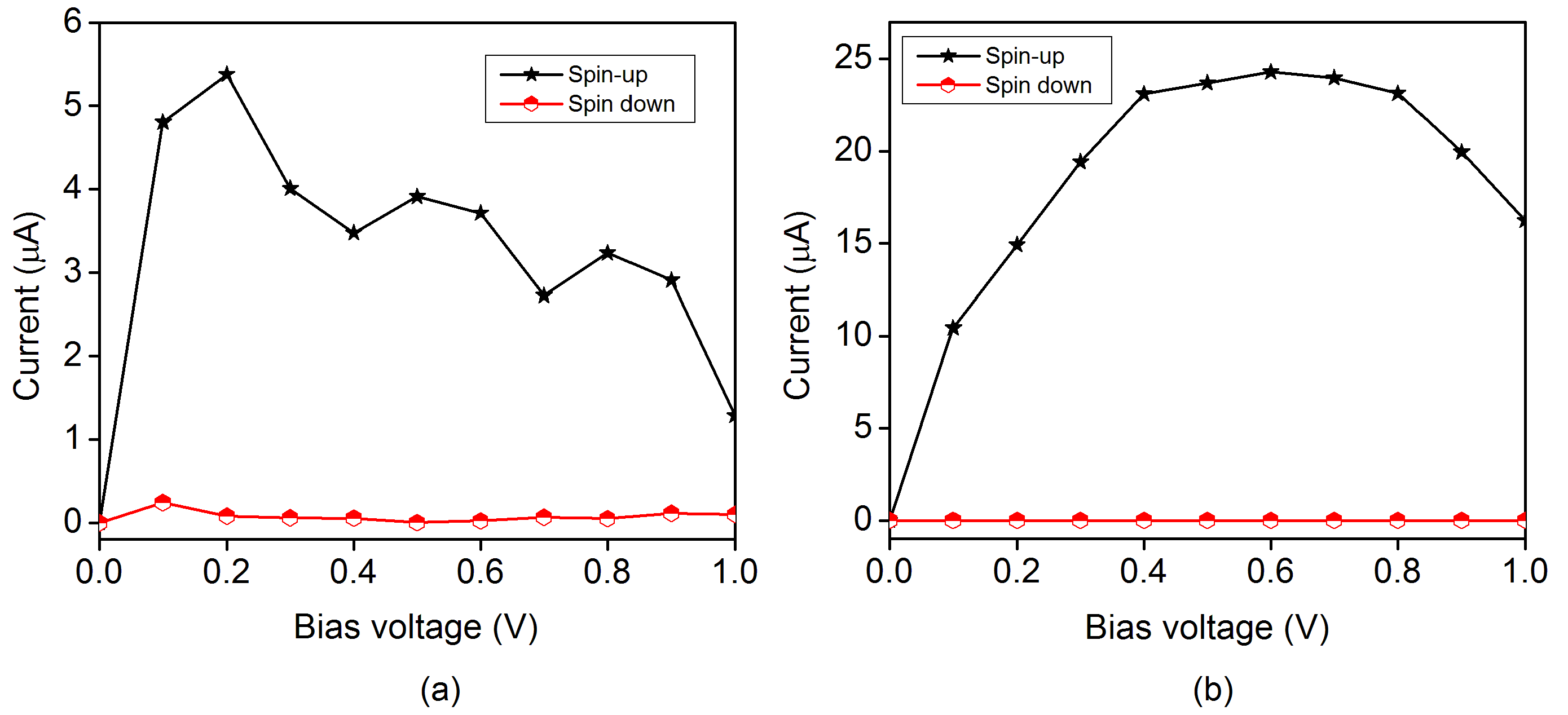}
\par\end{centering}
\caption{\label{fig:spin-IV}The calculated spin oriented I-V characteristic
of CO passivated ZGaN for (a) CO-ZGaNNR-CO and (b) CO-ZGaNNR-H nanoribbons.}

\end{figure}

To further confirm the half metallic behavior in selected structures
as observed in the band structures {[}Fig. \ref{fig:All-BS} (a),
(b){]}, we computed spin polarized I-V characteristics. Fig \ref{fig:spin-IV}
depicts the spin dependent currents for CO-ZGaN-CO and CO-ZGaN-H structures.
It is clearly visible in both of these images that the entire current
conduction takes place only due to spin up (majority spin) electrons.
As the current due to spin down (minority spin) electrons remains
essentially zero, it confirms the observed HM property of these structures. 

\subsection*{Spin filtering efficiency}

\begin{figure}[H]
\begin{centering}
\includegraphics[scale=0.3]{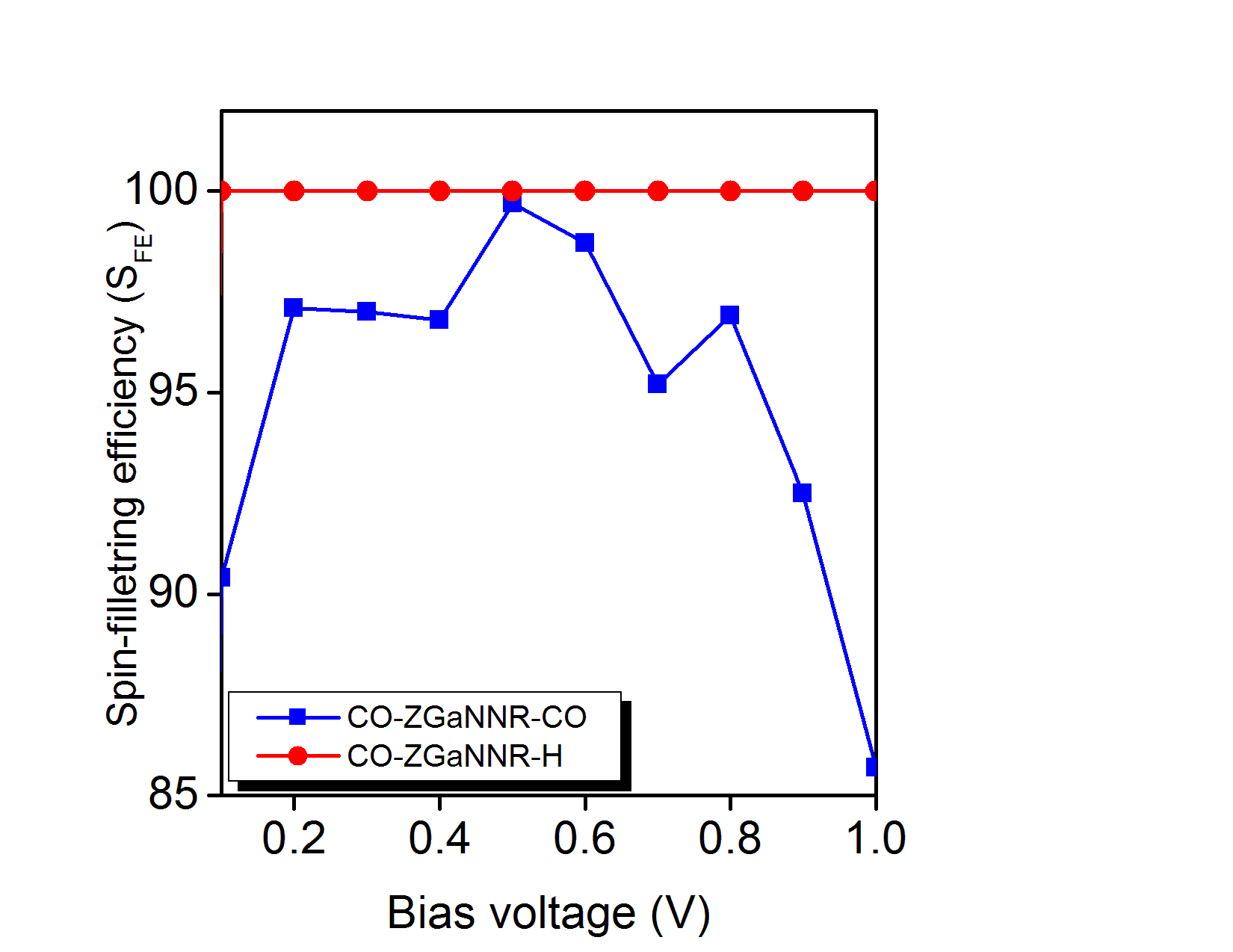}
\par\end{centering}
\caption{\label{fig:spin-filtering}The calculated spin filtering efficiencies
of CO-ZGaNNR-CO and CO- ZGaNNR-H.}
\end{figure}

The spin I-V characteristics of considered structures exhibit interesting
features. We find that for a given voltage, the magnitude of the current
due to one spin orientation is much more than that due to the other.
This imbalance between the current contributed by the two different
spin orientations is an important parameter for spin filtering device
applications. The ability of a material or a device to select a particular
spin direction can be quantified in terms of spin filtering efficiency
($S_{FE}$) defined as \cite{caffrey2013,kharadi2020,chen2018}:

\begin{equation}
S_{FE}=\frac{I_{spin\uparrow}-I_{spin\downarrow}}{I_{spin\uparrow}+I_{spin\downarrow}}\times100\%
\end{equation}

where, $I_{spin\uparrow}$ and $I_{spin\downarrow}$ are the magnitudes
of the currents for spin-up and spin-down electrons, respectively.
Fig \ref{fig:spin-filtering} shows the behavior of spin filtering
efficiency as a function of applied voltage for CO-ZGaNNR-CO and CO-ZGaNNR-H
structures. With respect to the applied bias voltage, we note the
following trends: (a) for CO-ZGaNNR-CO, $S_{FE}$ varies in the range
86$\%$-100$\%$, while (b) for CO-ZGaNNR-H, $S_{FE}$ stays constant
at 100$\%$. This behavior is fully consistent with the variation
of the currents of the two spin orientations, with the applied bias
voltage, for the two types of nanoribbons (see Fig \ref{fig:spin-IV}).
These results suggest that if one is looking for a perfect spin filter,
CO-ZGaNNR-H is a strong candidate for the purpose.

\subsection*{Different CO concentration}

\begin{figure}[H]
\begin{centering}
\includegraphics[scale=0.35]{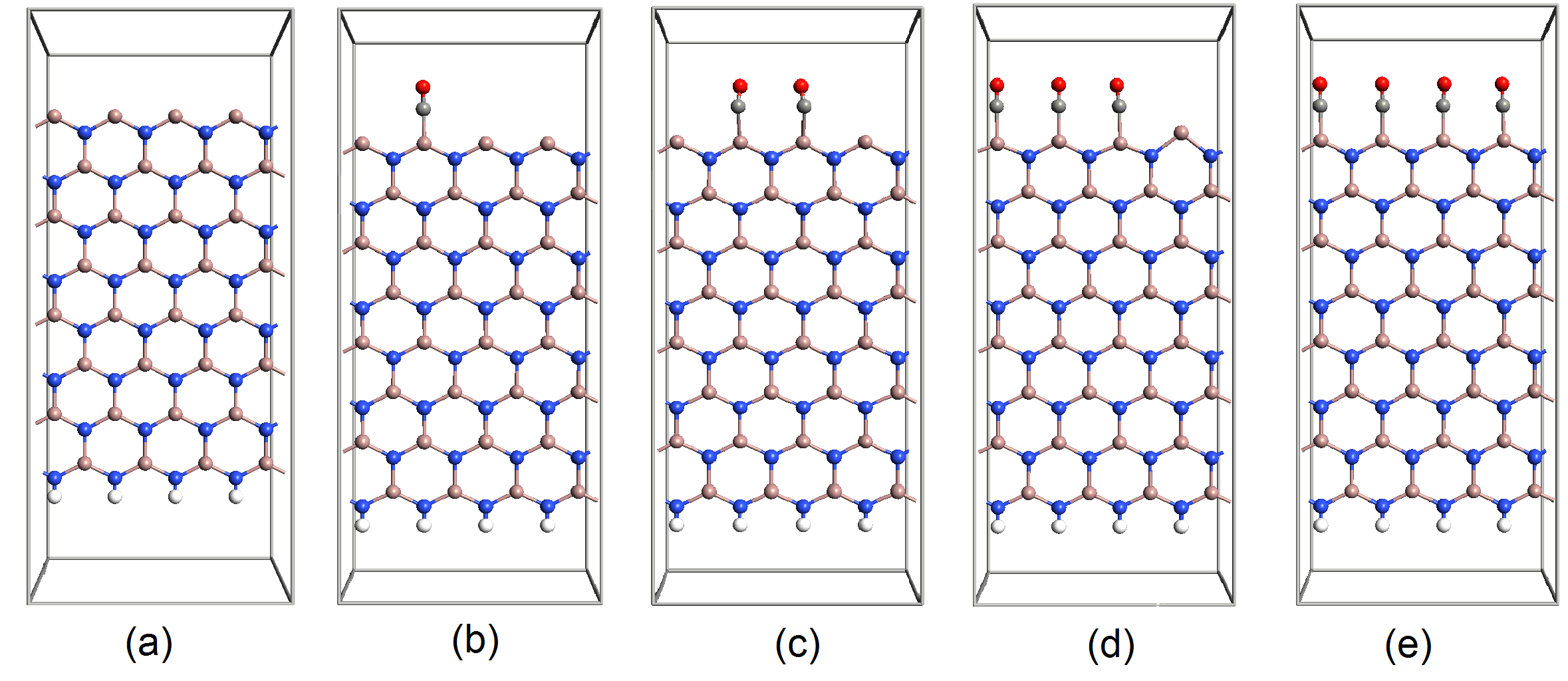}
\par\end{centering}
\caption{\label{fig:Conce-config}The Schematic diagrams of different concentration
of CO on most energetically favorable configuration CO-ZGaN-H for
(a) 0\%, (b) 25\%, (c) 50\%, (d) 75\% and (e) 100\% at width-8 and
4 repetition.}

\end{figure}

In the present work we have also study the different CO concentration
effect on the of most energetically favorable configuration (CO-ZGaN-H).
Different edge coverage (0$\%$ to 100$\%$ with interval of 25$\%$,
where 0$\%$ means absence of CO and 100$\%$ means all edges are
passivated by CO molecule.) of CO is obtained for the analysis of
the electronic properties of CO-ZGaN-H (Fig \ref{fig:Conce-config}).

\begin{figure}[H]
\begin{centering}
\includegraphics[scale=0.2]{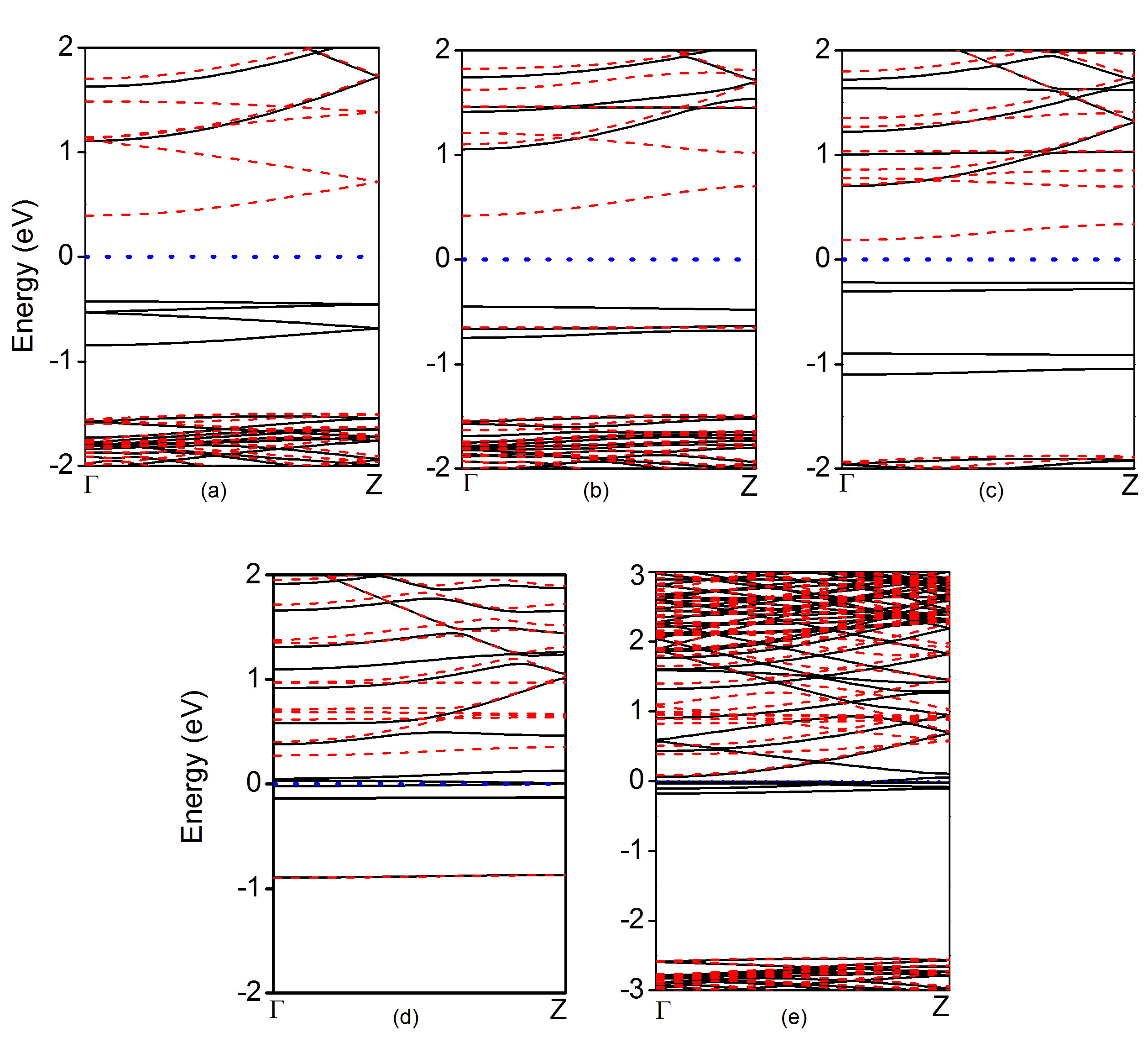}
\par\end{centering}
\caption{\label{fig:Conce-BS}The calculated electronic band structure of different
concentration of CO on most energetically favorable configuration
CO-ZGaN-H for (a) 0\%, (b) 25\%, (c) 50\%, (d) 75\% and (e) 100\%
at width-8 and 4 repetition.}

\end{figure}

It is noticed that the different edge coverage of CO profoundly alter
the electronic properties of considered structures. The band structures
of CO-ZGaN-H with different CO concentrations is illustrated in Fig
\ref{fig:Conce-BS}. Interestingly, it is noticed that as the edge
coverage of CO increases, the band gap decreases. The magnitude of
obtained band gap is 1.5 eV (spin-up) $\&$ 1.9 eV (spin-down), 1.5
eV (spin-up) $\&$ 1.1 eV (spin-down), 0.9 eV (spin-up) $\&$ 2.1
eV (spin-down), metallic (spin-up) $\&$ 1.1 eV (spin-down), and metallic
eV (spin-up) $\&$ 2.6 eV (spin-down) for 0$\%$, 25$\%$ and 50$\%$
CO coverage. In contrast, for 75$\%$ and 100$\%$ CO coverage, the
half metallic behavior is obtained as already discussed in the previous
section. Thus, variation in the CO coverage could be a potential way
to tailor the electronic band gap of ZGaNNR.

\section*{Conclusion}

In summary, our DFT based first-principles calculations demonstrate
that after passivation of CO molecule on one or both the edges of
ZGaNNR, the nanoribbons exhibit metallic to half metallic character.
Of the two possible CO-passivated configurations, we find that the
nanoribbons passivated with CO on one edge (CO-ZGaNNR-H) are not only
more stable as compared to the ones passivated with it on both the
sides (CO-ZGaNNR-CO), but also they exhibit better spin-filtering
efficiency. Furthermore, we also observe variations in Fermi energy
as a function of passivation, suggesting that the edge passivation
can also be used to achieve doping in these systems. The negative
differential resistance observed in these structures also suggests
their possible applications in fabricating oscillators and amplifiers.
Therefore, we believe that structures can be used to fabricate a variety
of devices such as CO sensors, spin filters, oscillators, and amplifiers. 

\bibliographystyle{unsrt}
\bibliography{Reference-CO}

\end{document}